\documentclass[amsart,preprint,unsortedaddress]{revtex4-1}
\pdfoutput=1
\usepackage{geometry} 
\usepackage{enumerate}
\usepackage{amsmath}
\usepackage{amssymb}
\usepackage[dvips]{graphicx}
\usepackage{color}
\usepackage{listings}
\usepackage{tabularx}
\usepackage{gensymb}
\usepackage[]{algorithm2e}
\usepackage[footnotesize]{subfigure}
\usepackage{float}

\begin{document}
\title{Arbitrarily accurate representation of atomistic dynamics via Markov Renewal Processes}
\author{Animesh Agarwal}
\affiliation{Theoretical Biology and Biophysics, Los Alamos National Laboratory, New Mexico 87545, United States}
\author{Sandrasegaram Gnanakaran}
\affiliation{Theoretical Biology and Biophysics, Los Alamos National Laboratory, New Mexico 87545, United States}
\author{Nicholas Hengartner}
\affiliation{Theoretical Biology and Biophysics, Los Alamos National Laboratory, New Mexico 87545, United States}
\author{Arthur F. Voter}
\affiliation{Physics and Chemistry of Materials, Los Alamos National Laboratory, New Mexico 87545, United States}

\author{Danny Perez}
\email{danny\_perez@lanl.gov }
\affiliation{Physics and Chemistry of Materials, Los Alamos National Laboratory, New Mexico 87545, United States}

\begin{abstract}

Atomistic simulations with methods such as molecular dynamics are extremely powerful tools to understand nanoscale dynamical behavior. The resulting trajectories, by the virtue of being embedded in a high-dimensional configuration space, can however be difficult to analyze and interpret. This makes low-dimensional representations, especially in terms of discrete jump processes, extremely valuable. This simplicity however usually comes at the cost of accuracy, as tractable representations often entail simplifying assumptions that are not guaranteed to be realized in practice. In this paper, we describe a discretization scheme for continuous trajectories that enables an arbitrarily accurate representation in terms of a Markov Renewal Process over a discrete state space. The accuracy of the model converges exponentially fast as a function of a continuous parameter that has the interpretation of a local correlation time of the dynamics.

\end{abstract}
\maketitle

\section{Introduction}

Atomistic simulations using molecular dynamics (MD) are widely recognized as extremely powerful tools to investigate the behavior of complex dynamical systems such as materials and biomolecules. While the predictive power of MD is extremely high, the complexity of atomistic trajectories in the full phase space can be such that interpreting and rationalizing the results poses a significant challenge.
This has made the development of compact and interpretable representations of complex dynamics an area of intense interest over the past few decades \cite{torda1994algorithms,ma2005automatic,das2006low,tribello2012using,NADLER2006113,bowman2013introduction,sittel2014principal}.

In this paper, we consider coarse models that are defined on a discrete state space. This type of representation is particularly interpretable because it describes the dynamics in terms
of a sequence of waiting times punctuated by discrete jumps between states, i.e., in terms of a jump process. Developing a jump process that captures the dynamics of a continuous system is a two step process that involves  i) developping
a discretization procedure that maps a continuous trajectory $x(t)$ into a sequence of discrete states and transition times ($\{X_n, T_n\}$), and ii) constructing the conditional jump probability distribution $Pr(T_{n+1}-T_n<t, X_{n+1}=j | (X_0, T_0), (X_1, T_1),...,(X_n=i, T_n))$ that statistically reproduces the dynamics of the discretized process.
For example, a popular approach begins by introducing a number of sets in configuration/phase space (which can be taken to tesselate that space, for simplicity) and
by assigning to each point $x(t)$ the discrete index $X$ of the set that contains $x(t)$. One then assumes that the
state-to-state (i.e., set-to-set) dynamics is Markovian, i.e.,  $Pr(T_{n+1}-T_n<t, X_{n+1}=j | (X_0, T_0), (X_1, T_1),...,(X_n=i, T_n))=Pr(X_{n+1}=j | X_n=i)(1-\exp(-l_i t))=p_{ij}F_i(t)$, where $l_i$ is the total escape rate out of set $i$. Note that in the Markovian case, the distribution of escape times is independent of the final state $j$. This assumption gives rise to a representation in terms of a continuous-time Markov chain (CTMC). CTMC are extremely compact, can be easily analyzed formally, and allow for the efficient generation of new discrete trajectories using the celebrated BKL \cite{bortz1975new} or Stochastic Simulation algorithms \cite{gillespie1976general}. These favorable characteristics have made CTMC representations extremely popular in the computational sciences \cite{voter2007introduction,noe2007hierarchical,chodera2007automatic,bowman2013introduction}.

For all of their powerful features, CTMC representations of atomistic dynamics are however generally not exact because the mapping from a continuous to a discrete state-space is not guaranteed to preserve the Markov property. The mapping between a continuous and a discrete Markovian dynamics has been recently revisited using the theory of quasi-stationary distributions (QSD) \cite{le2012mathematical,lelievre2018mathematical,di2019exit}. This analysis has shown that the accuracy of the Markovian representation can be quantified based on the spectral properties of the generators of the dynamics restricted to each set by absorbing boundary conditions. This representation can be shown to becomes exact in the limit where the spectral gap of each set becomes infinitely large \cite{le2012mathematical}, so that the relaxation time within each set tends to zero, which is in general not the case for finite sets \cite{suarez2016accurate}. That being said, in practice, if sufficiently metastable sets can be defined such that relaxation within every single set is fast on timescales characteristic of escapes, the corresponding Markovian representation will be very accurate.

The limitation of CTMCs stems from the assumption that the state-to-state dynamics is strictly memory-less. One possible solution is to invoke a more general form of the transition probabilities that includes a memory of the past \cite{suarez2016accurate,doi:10.1002/pro.2738}.
In this paper, we explore a simpler alternative where the dynamics is represented in terms of Markov Renewal Processes (MRPs), where Markovian constraints on the form of the probability distributions are slightly relaxed to
$Pr(T_{n+1}-T_n<t, X_{n+1}=j | X_n=i )=p_{ij}F_{ij}(t)$; i.e., the difference with CTMCs is that the cumulative distribution of residence times $T_{n+1}-T_n$ given an initial state $X_n=i$ is now a general non-decreasing function of $t$ and of the final state $X_{n+1}=j$ such that $F_{ij}(0)=0$ and $\lim_{t\rightarrow \infty} F_{ij}(t)=1$ \cite{korolyuk1975semi}. Building on an analysis in terms of quasi-stationary distributions (QSDs), we show that a novel discretization scheme allows for an arbitrarily accurate representation via MRPs for any number and shape of sets. It can therefore be expected to be a powerful representation to accurately represent the dynamics of a wide range of systems.

\section{Derivation}
In the following, we consider the problem of obtaining a simplified representation of an overdamped Langevin dynamics in $R^N$ in terms of a jump process over a discrete state space. We are specifically interested in identifying a mapping from continuous dynamics to discrete jump processes that can provide an arbitrarily improvable approximation to the exact discretized dynamics for {\em any} number and shape of sets. We first consider such a jump process in discrete time, and then generalize to continuous time.
Note that the results derived below also hold in the context of the Langevin equation at finite friction. The mathematical analysis is however extremely involved \cite{nier2013boundary}, so this general case is not explicitly analyzed in the following. As we will show, it also holds for any process that converges to a unique QSD when confined by absorbing boundaries. The approach we propose here can therefore be expected to be useful in a very wide range of conditions, well beyond that of obtaining a reduced representation of atomistic dynamics.

\subsection{State Definition}
\label{subsec:states}
\begin{figure}
\centering     
\includegraphics[width=85mm]{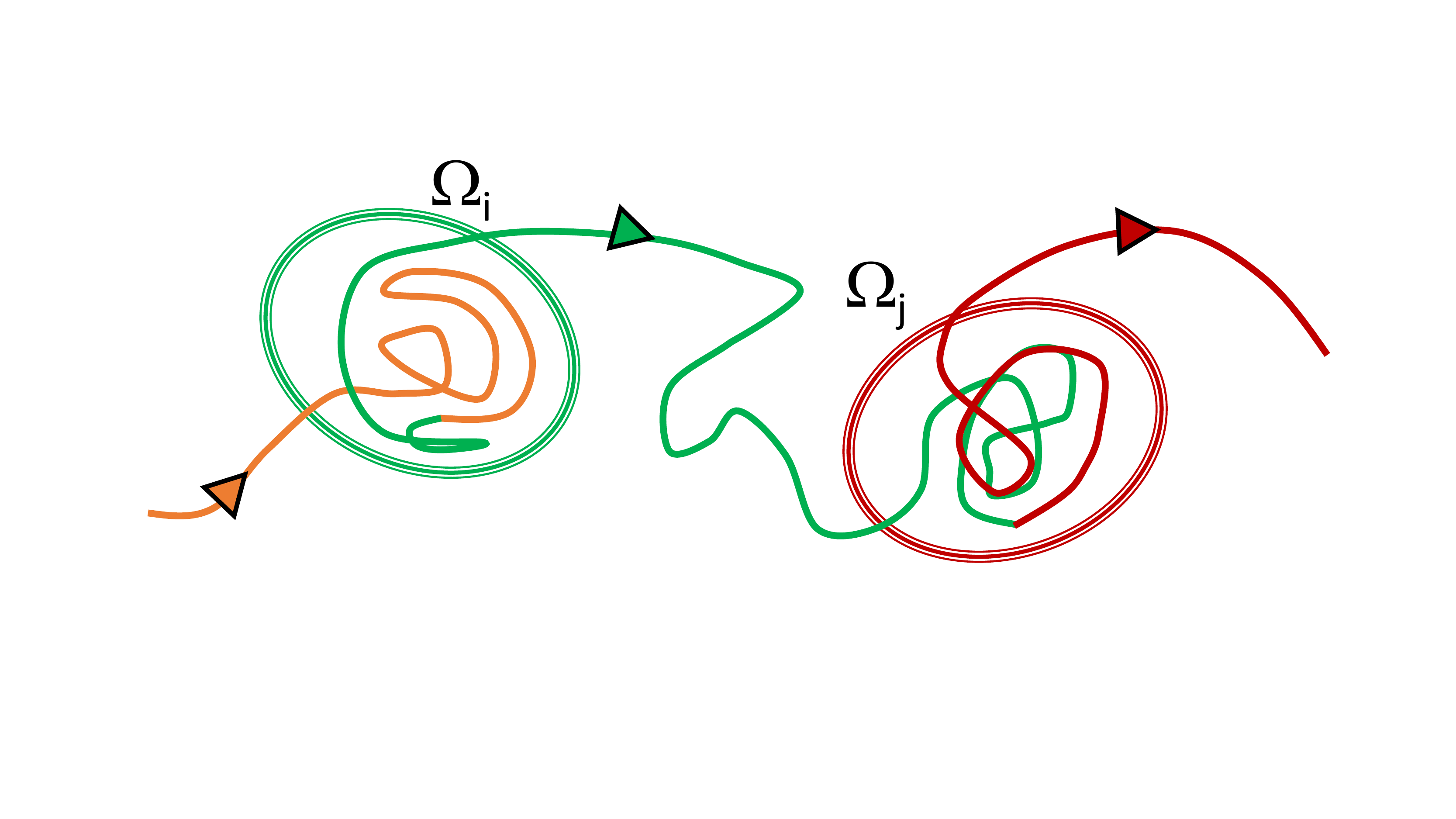}
\caption{Illustration of the trajectory discretization process. Two sets $\Omega_i$ and $\Omega_j$ are represented by ellipses and the trajectory $x(t)$ by a continuous line. Each point along the trajectory is assigned a discrete index that corresponds to the last set where it remained for a time $t_c$ without escaping, as indicated by its color.
}
\label{fig:cartoon}
\end{figure}

The first task is to define the discrete state space in which the jump process evolves. To do so, consider a number of arbitrarily disjoint connected sets $\Omega_i$ defined over the $N$-dimensional configuration space. In contrast to the common approach discussed above, we assign to each $x(t)$ an integer $X$ that indexes the set where the process last remained for at least a time $t_c$ without escaping; $t_c$ can be chosen to be different for each set, but it will in the following be taken as a global constant for simplicity of notation.
The index assigned to a given point $x(t^*)$ therefore corresponds to the last set where the trajectory $x(t\leq t^*)$ remained for at least $t_c$ without escaping. See Fig.\ \ref{fig:cartoon} for a schematic illustration of the mapping. The state definition is therefore not purely geometric, but depends on the past history of the process $x(t)$. As will be shown below, this representation becomes especially powerful when $t_c$ is chosen to be larger than a suitably-defined local "correlation" or "memory" time of the trajectory. Intuitively, this definition acts to limit the coupling between the consecutive jumps, which is key to making the representation local in time.

Indeed, the main result of this paper is that the statistics of trajectories discretized in this way can be arbitrarily approached by a MRP by increasing $t_c$, no matter the number and shapes of the sets $\{\Omega\}$ used in the discretization. We first consider the discrete-time case, which we then extend to continuous time. We finally use numerical simulation of biological systems to show that this representation can be very accurate even for relatively small values of $t_c$.

This definition is based on insights gained through a formal analysis \cite{le2012mathematical} of the Parallel Replica Dynamics \cite{voter1998parallel,perez2015parallel} and Parallel Trajectory Splicing \cite{perez2016long,agarwal2019computing} accelerated molecular dynamics method \cite{perez2009accelerated,zamora2018accelerated} that demonstrated that the statistics of trajectories conditioned on having remained for a long time within a set become especially simple.

\subsection{Discrete time jump process}

The first important result is that the discrete-time jump process that describes the sequences of states visited by the discretized process approaches a simple discrete-time Markov chain as $t_c  \rightarrow \infty$.
To show this, consider the evolution of an ensemble of trajectories initialized at some point $x_0 \in \Omega_{0}$ conditioned on remaining inside $\Omega_{0}$. This problem has been rigorously investigated in Ref.\ \onlinecite{le2012mathematical}; only a broad outline is reproduced here for completeness and details can be found in the original reference.
The (unnormalized) probability distribution in configuration space evolves according to a Fokker-Planck equation of the form:

$$\frac{\partial p}{\partial t}= (-\nabla V\cdot \nabla +\beta^{-1}\Delta)p $$ with
\begin{equation}
\begin{split}
  p(x,0) & =\delta(x_0) \\
  p(x,t) & = 0 \quad  \mathrm{on} \quad  \partial\Omega_{0},
\end{split}
\end{equation}
where $V(x)$ is the underlying potential energy landscape and $\beta$ is the inverse temperature.
This relation can be rewritten in the eigen-basis of the generator ($\{-\lambda_k,\phi_k\}$) and of its adjoint ($\{\psi_k\}$) as:
\begin{equation}
\begin{split}
p(x,t) & = \sum_k \exp(-\lambda_k t) \phi_k(x) \int_\Omega p_0(y,0) \psi_k(y) dy \\
       & =\sum_k \exp(-\lambda_k t) \phi_k(x) \psi_k(x_0)
\end{split}
\end{equation}
where the eigenvalues are ordered such that $0<\lambda_1<\lambda_2\leq\lambda_3,...\ $, where we note that, for this dynamics, $\lambda_1$ is guaranteed to be non-degenerate \cite{le2012mathematical}.

Taking the limit $t$ $\rightarrow$ $\infty$, the last equation becomes:
$$ \lim_{t\rightarrow \infty} p(x,t) =  \exp(-\lambda_1 t)  \phi_1(x) \psi_1(x_0) + \mathcal{O}(\exp(-\lambda_2 t)),  $$

or, in normalized form,
\begin{equation}
  \label{eq:qsd_lim}
 \lim_{t\rightarrow \infty} \tilde{p}(x,t) =  \frac{p(x,t)}{\int_{\Omega_0} p(x,t) \ dx}  =  \phi_1(x) + \mathcal{O}(\exp(-(\lambda_2-\lambda_1) t)).
\end{equation}

Conditional on not having escaped, the normalized distribution function inside $\Omega_{0}$ tends to $\phi_1(x)$, the so-called quasi-stationary distribution (QSD) of $\Omega_{0}$ for all $x_0$, up to terms that decay exponentially with $t$. What are then the first escape statistics out of $\Omega_{0}$ after having spend a time $t\rightarrow \infty$ within it? The previous elementary derivation directly implies an important result: the escape flux out of a given infinitesimal surface element centered on point $x_s$ on $\partial\Omega_{0}$ --- which is proportional to $\nabla \phi_1(x_s) \cdot n(x_s)$, where $n(x_s)$ is the local normal to $\partial \Omega_{0}$ --- is also independent of $x_0$ and of the escape time, since the probability density tends to a unique distribution $\phi_1(x)$ for all initial conditions $x_0$.
See Proposition 3 in Ref.\ \onlinecite{le2012mathematical} for a rigorous derivation.

With this result, return to the issue at hand: consider a long continuous trajectory $x(t)$ that enters set $\Omega_i$ at $t=0$ and then remains within it for a time $t_c\rightarrow \infty$ without escaping, at which time the discrete trajectory made a transition to state $X_n=i$. What is then the probability that the next set in which the trajectory spends $t_c$ is $\Omega_j$ (and hence that the discrete trajectory next makes a transition to $X_{n+1}=j$)?
From the previous derivation, the distribution of first escape points on $\partial \Omega_i$, and hence of the future of the trajectory after it escapes, is independent of $x(0)$ and of the time at which the escape ultimately occurs. It follows that this distribution is also independent of $x(t<0)$ and hence of $\{X_{k<n}\}$.

This entails the first important result of this paper:

In the limit $t_c$ $\rightarrow$ $\infty$, the probability that a discretized trajectory currently in state $i$ next makes a transition to state $j$ is independent of the residence time in $i$ and of the states visited before $i$, i.e., it is of the form $Pr(X_{n+1}=j | X_n=i)=p_{ij}$ so that the corresponding discrete-time jump process is Markovian. Note that as $t_c$ $\rightarrow$ $\infty$, $p_{ij} \rightarrow p_j$. As this result will be shown below to also be useful in the finite $t_c$ limit, the dependence on $i$ is here kept explicit.

A few comments are in order. First, this result holds no matter how the sets $\Omega_i$ are defined. As will be shown below, the rate of convergence with respect to $t_c$ does depend on the set definition, but eventual convergence to an arbitrary precision does not. Second, this result is not exclusive to overdamped dynamics, or even to ordinary Langevin dynamics. Indeed, this result only relies on eventual convergence to a unique QSD in each set, conditional on not escaping the set. While proving existence of a unique QSD can be in general difficult, a large number of killed processes are known to converge to a unique QSD when they are conditioned on surviving for an arbitrary long time \cite{collet2012quasi}.

\subsection{Continuous-time jump process}

We now investigate the corresponding jump dynamics in continuous time. As shown in Ref.\ \onlinecite{le2012mathematical}, strong statements can be made of the distribution of residence time in {\em set} $\Omega_i$ given that $x(t)$ remained within it for $t_c \rightarrow \infty$ without escaping. Most notably, the distribution of exit times is exponential and escape points and times are uncorrelated random variables (c.f. Proposition 3 in Ref.\  \onlinecite{le2012mathematical}). However, the same cannot be said of the corresponding distribution of residence time in {\em state} $i$ except for the fact that it also cannot
depend on $x(t<0)$ and hence on $\{X_{k<n}\}$. 
Indeed, while the escape time from the initial set $\Omega_i$ will become independent of the entry time and entry point as $t_c \rightarrow \infty$, the time at which the trajectory first spends a time $t_c$ in another set $\Omega_j$ can depend on $j$, For example, for final sets that are distant from $\Omega_i$, the transit time from $\Omega_i$ to $\Omega_j$, introducing a final-state dependent contribution that does not have to be exponential. Similarly, if $\Omega_j$ is weakly metastable, the trajectory might have to enter and exit the set $\Omega_j$ many times before spending $t_c$ entirely within $\Omega_j$, again introducing a final-state dependent contribution.

This directly entails the second main result of this paper (again in the limit $t_c \rightarrow \infty$):

$Pr(T_{n+1}-T_n<t, X_{n+1}=j | X_n=i)$,
the probability that the next transition from state $i$ will be to state $j$ and will occur before a time $t$ has elapsed since the last transition into state $i$, is of the form $p_{ij}F_{ij}(t)$, where $F_{ij}$ is a non-decreasing function such that $F_{ij}(t<t_c)=0$ and $\lim_{t\rightarrow \infty}  F_{ij}(t)=1$,
i.e., the continuous-time jump process that describes the sequence of visited states and residence times becomes a Markov Renewal Process \cite{korolyuk1975semi}.
As before, the result holds for any definition of the sets {$\{\Omega$\}}.
Further, it again not only holds for overdamped dynamics, as it only relies on the existence of a unique QSD within each set, which is known to exist for a large number of processes \cite{collet2012quasi}.

\section{Discussion}

The previous result is strictly exact only in the limit where $t_c$ tends to infinity. In this limit, the jump process is also unfortunately non-informative, as the residence time within each state diverges, hence providing very little information on whereabouts of the system between jumps. Key to the practical relevance of this representation is the fact that convergence is exponentially fast as a function of $t_c$. It was indeed shown in Prop. 6 of Ref. \onlinecite{le2012mathematical} that the difference between the joint distributions of first escape times and first escape points conditioned on remaining in set $\Omega_i$ for an infinite and a finite amount of time $t_c$, respectively, decays as $\exp(-(\lambda_2-\lambda_1)t_c)$, as suggested by Eq.\ \ref{eq:qsd_lim}. Note that this holds for any functions of the escape times and points out of $\Omega_i$.
Hence, local deviations between the MRP representation and the actual statistics of the discretized Langevin process decay exponentially with $t_c$, on a state-per-state basis. Convergence is especially fast when the sets can be chosen to be sufficiently metastable, i.e.,
to have a  sufficiently large spectral gap $\lambda_2-\lambda_1$. Then, $t_c$ can be chosen so as to guarantee an accurate representation ($t_c > 1/(\lambda_2-\lambda_1)$), while still preserving a good time resolution ($t_c < 1/\lambda_1$). Note that it was recently shown that exponential convergence to the QSD also occurs for Langevin dynamics at finite friction \cite{nier2013boundary} and for a number of other random processes \cite{champagnat2016exponential}. However, exponential convergence is not guaranteed for arbitrary dynamics, so slow convergence of the MRP representation with increasing $t_c$ is possible in some cases.

An important advantage of the mapping to MRPs is that the accuracy of a jump process based on sub-optimal set definitions can nonetheless be made arbitrarily high by simply increasing $t_c$. The MRP representation can therefore be expected to be useful in a wide range of conditions, even with a rather naive definition of the sets  $\{{\Omega}\}$, which we demonstrate using numerical examples below.  It should however be kept in mind that if sufficiently metastable sets cannot be defined, the trade-off between resolution and accuracy could be unfavorable,  resulting in either an accurate yet uninformative representation or an inaccurate but informative one.

\section{Numerical Results}

In order to demonstrate the accuracy of the MRP representation, we concentrate on $P_{ij}(t)$, the probability that the system is found in state $j$ at time $t$ given that it entered state $i$ at time $0$, i.e.,:

\begin{equation}
P_{ij}(t) = Pr\{X(t)=j | X(0)=i, T_0=0\}, i, j \in X; t \geq 0.
\end{equation}

For MRPs, $P_{ij}(t)$ is given by the solution of so-called renewal equations:

\begin{equation} \label{renewal}
P_{ij}(t) = (1-\sum_{j \in \{X\}} p_{ij}F_{ij}(t))\delta_{ij} + \sum_{k\in \{X\}} (p_{ik}F_{ik}\ast P_{kj})(t)
\end{equation}
where the second term on the R.H.S. is a convolution:
\begin{equation}
(A\ast B)(t) = \int_{0}^{t} A(t-s) dB(s)
\end{equation}
so that Eq. \ref{renewal} can be rewritten as:
\begin{equation}
P_{ij}(t) = (1-H_{i}(t))\delta_{ij} + \sum_{k \in \{X\}} \int_{0}^{t} P_{kj}(t-s) p_{ik}F_{ik}(ds)  ; t \geq 0; i \in \{X\}
\end{equation}

In the following, renewal equations were numerically integrated using the algorithm proposed in Ref.\ \onlinecite{tortorella1990closed}.

We calculate the state-to-state transition probabilities in two well studied biomolecular systems: alanine dipeptide and chignolin. We generated 14 1-$\mu$s long trajectories of alanine dipeptide where snapshots were saved every 2 ps. The technical details of these simulations are reported in the supplementary material. The 1-$\mu$s long Chignolin trajectory (where snapshots were saved every 0.2 ns) was provided by D.E. Shaw Research; technical details of the simulations are provided in Ref. \cite{lindorff2011fast}. For both systems, we employ Markov State Models (MSMs) \cite{bowman2013introduction} and Perron Cluster-Cluster analysis (PCCA) \cite{deuflhard2005robust} to define metastable sets. Details on the simulations and construction of the sets are provided in the supplementary material. Note however that, for the present purpose, these details are not essential as we will show that a high accuracy MRP model can be obtained even for naive set definitions.


\begin{figure}
\centering     
\subfigure[$\tau_c$=2 ps]{\label{ala-opt-4:a}\includegraphics[width=65mm]{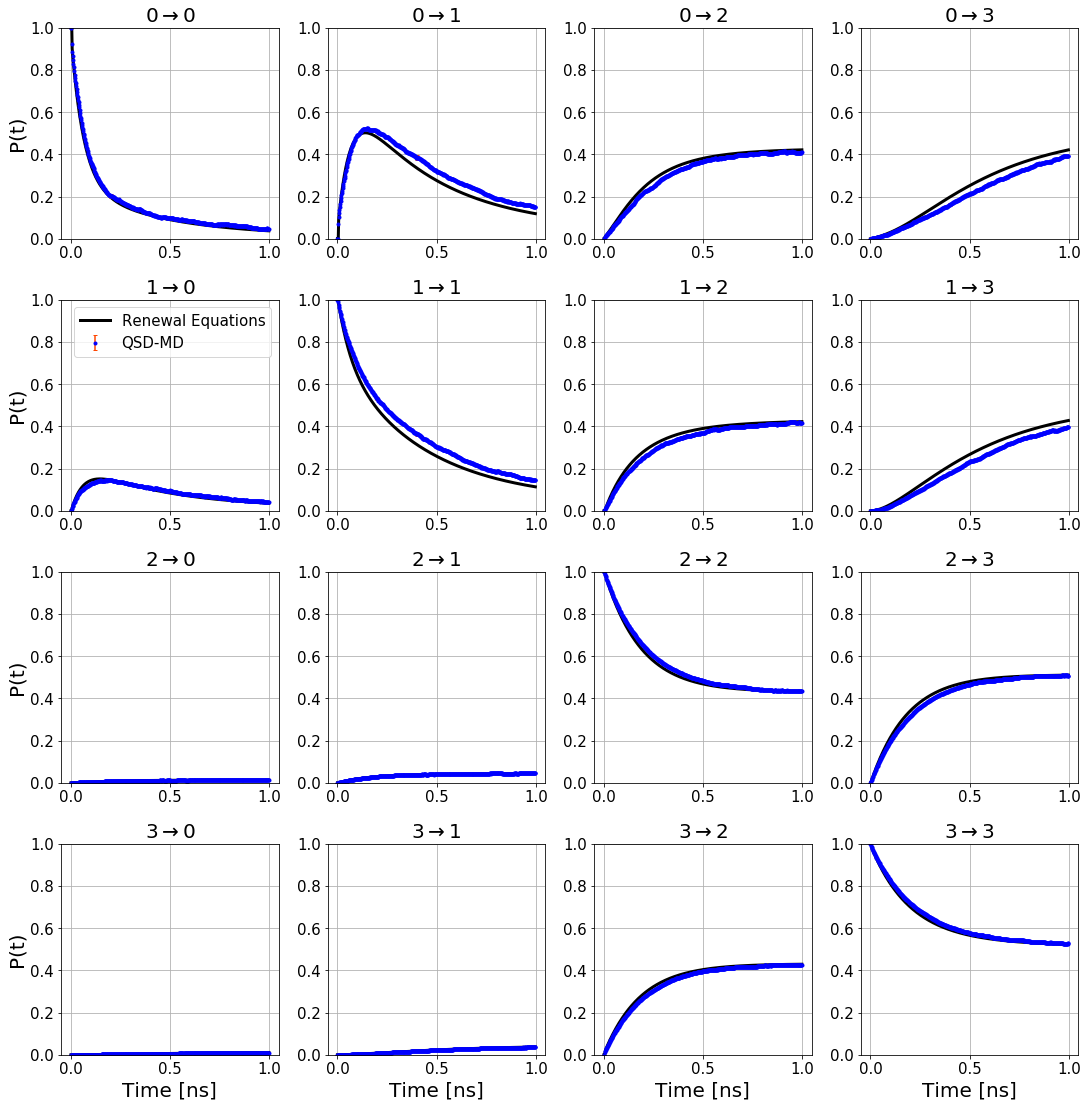}}
\subfigure[$\tau_c$=20 ps]{\label{ala-opt-4:b}\includegraphics[width=65mm]{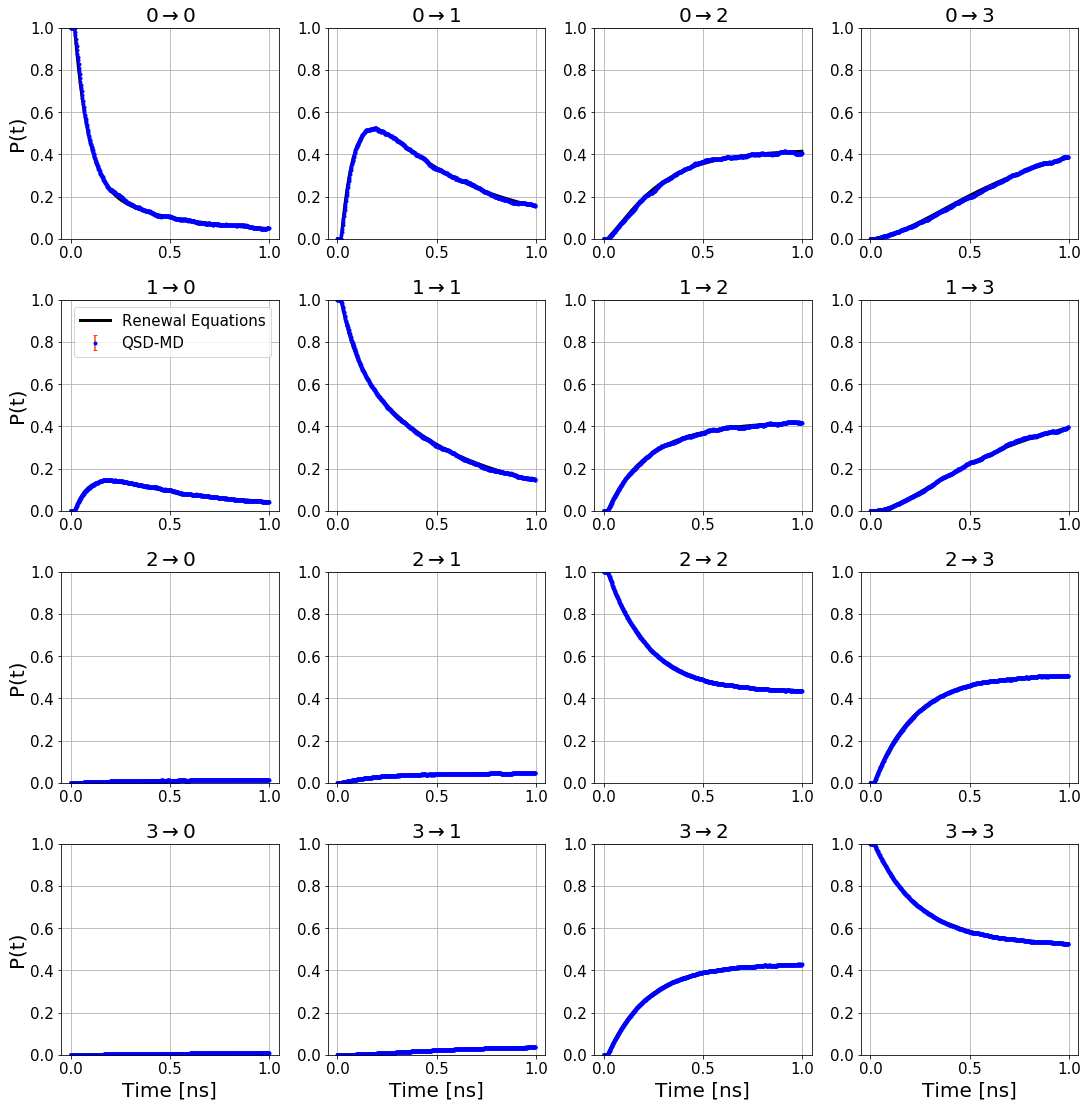}}
\subfigure[$\tau_c$=40 ps]{\label{ala-opt-4:c}\includegraphics[width=65mm]{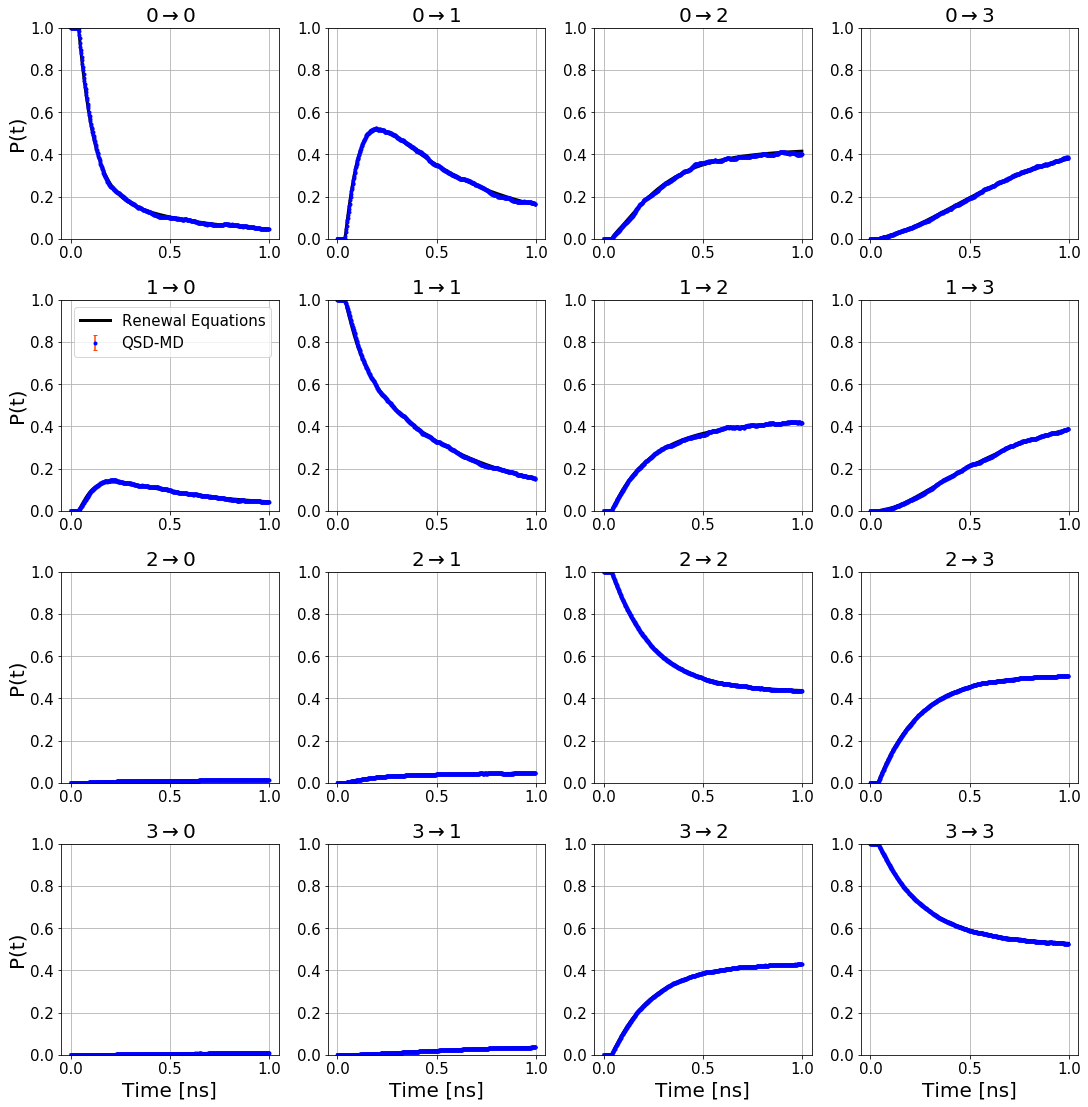}}
\caption{Transition probabilities computed from the discretized trajectory and the MRP model between the four PCCA
states of alanine dipeptide.}
\label{ala-opt-4}
\end{figure}

With the set boundaries defined, we discretized the continuous trajectory following the procedure discussed in Section \ref{subsec:states}, namely each point $x(t)$ was assigned a discrete state $X$ according to the set where the trajectory last remained for least a time $t_c$ without escaping. In order to assess the accuracy of the MRP, we compare the transition probabilities directly computed from the discretized trajectory (the gold standard in this context), to those obtained by solving the MRP using renewal equations. These equations were parametrized by fitting to the observed transition time distributions $p_{ij}F_{ij}(t)$ between every pair of states again using the discretized trajectory. Fig.\ \ref{ala-opt-4} shows the two sets of transition probabilities for alanine dipeptide for three different correlation times, $t_c =$ 2 ps, 20 ps and 40 ps, respectively. It is worth noting that the reference results also depend on $t_c$, and thus that both the target and the approximation change with that parameter.
It can be seen that even for the shortest correlation time (2 ps), the transition probabilities obtained by solving the renewal equations are in very good agreement with the directly measured probabilities, but small deviations can be observed. As expected, the results become indistinguishable from the direct reference at the larger values of $t_c$. In this case, the results show that the dynamics can be very well approximated by a MRP even for short correlation times, pointing to the fast convergence of the model with respect to $t_c$.

\begin{figure}
\centering     
\subfigure[$\tau_c$=2 ps]{\label{ala-ran-4:a}\includegraphics[width=65mm]{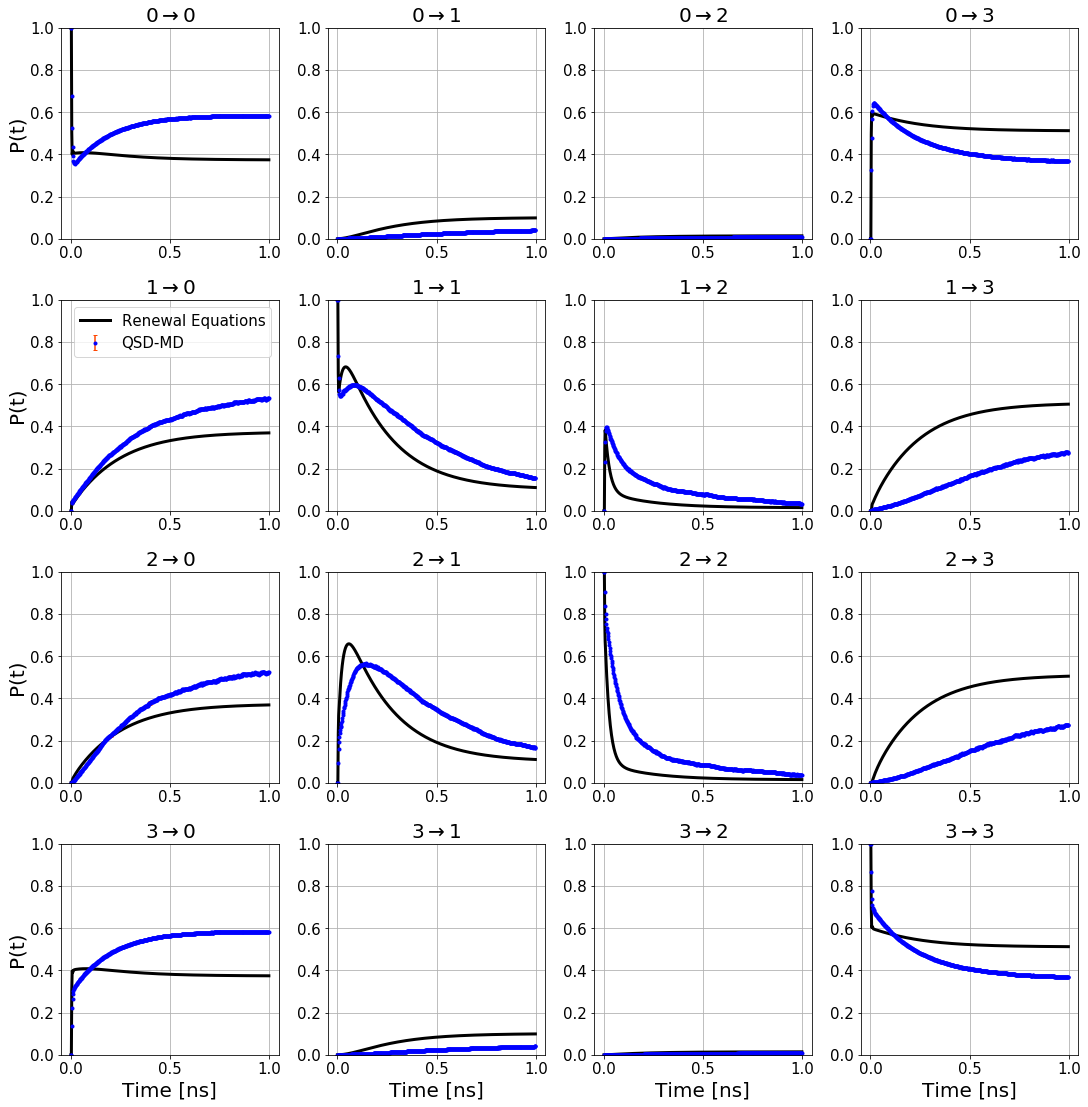}}
\subfigure[$\tau_c$=20 ps]{\label{ala-ran-4:b}\includegraphics[width=65mm]{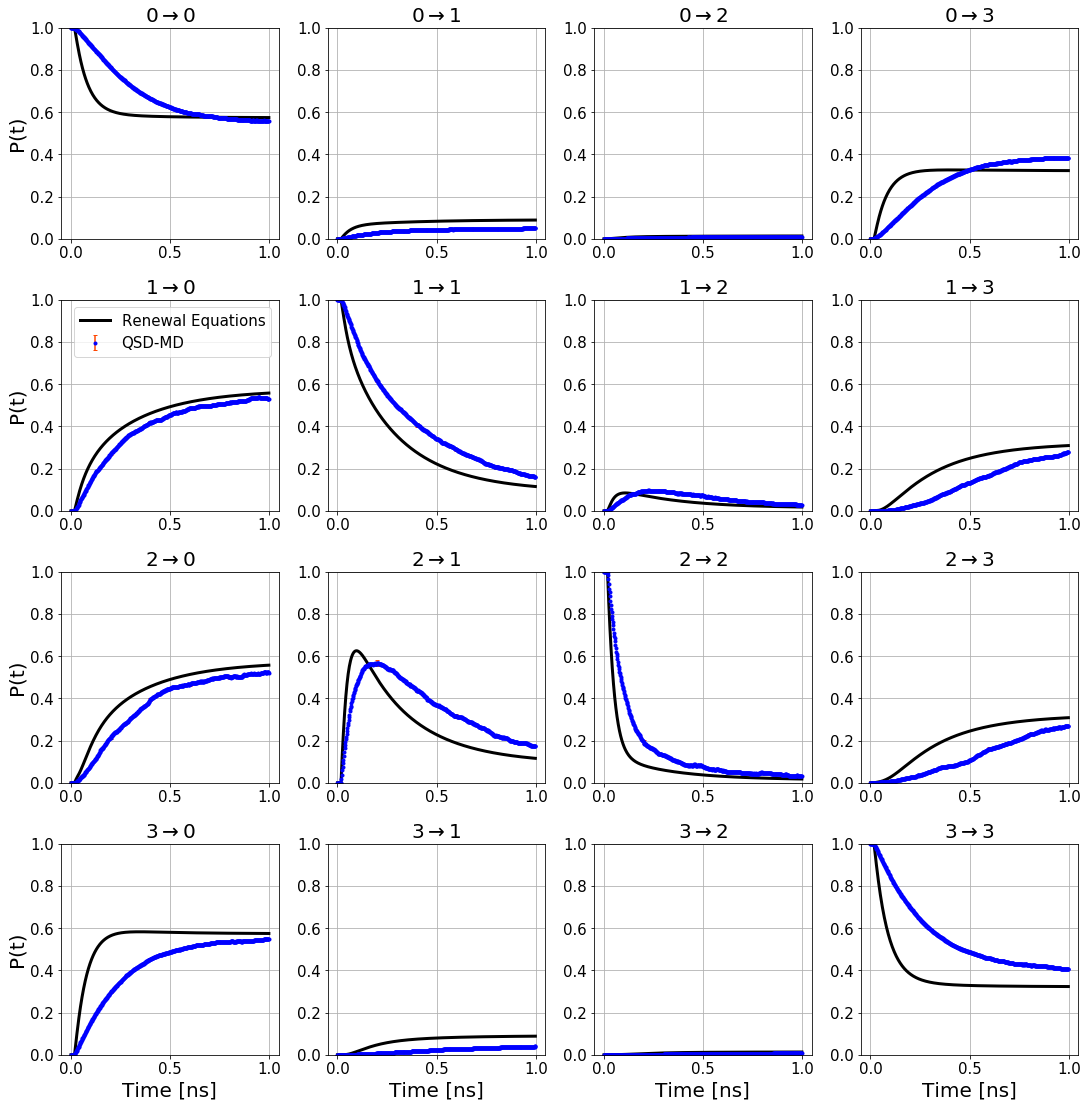}}
\subfigure[$\tau_c$=40 ps]{\label{ala-ran-4:c}\includegraphics[width=65mm]{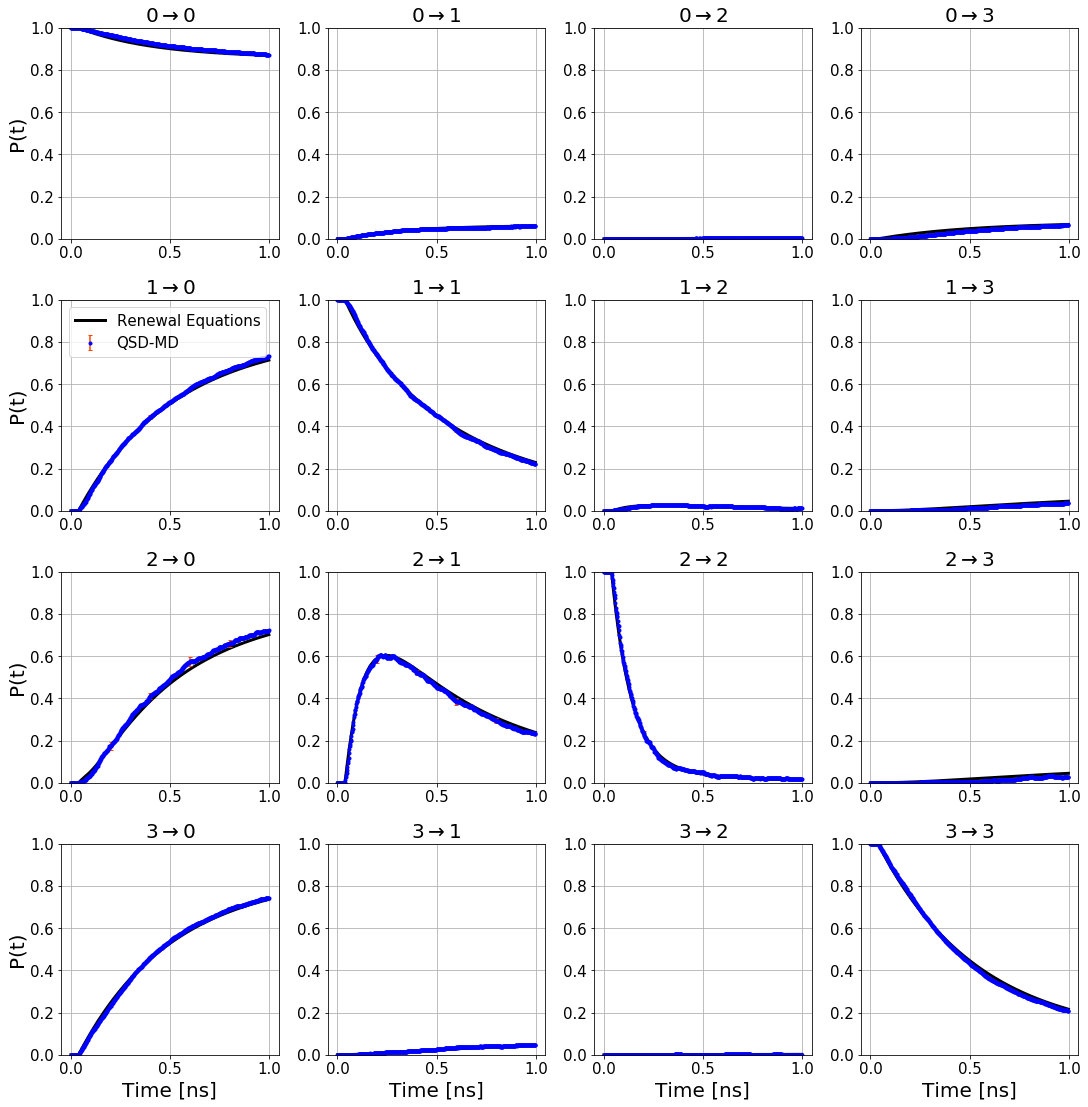}}
\caption{Transition probabilities computed from the discretized trajectory and the MRP  between the four rectangular (unphysical)
states of alanine dipeptide.}
\label{ala-ran-4}
\end{figure}

This convergence can be further demonstrated by the following experiment: instead of carefully defining sets so that they are as metastable as possible (and hence for which a Markovian approximation might be quite accurate), we instead arbitrarily construct 4 states by partitioning the $\phi-\psi$ space into four rectangular cells using  $\phi=0^{\circ}$, $180^{\circ}$  and $\psi= 0^{\circ}$, $180^{\circ}$ as dividing lines. We then discretize the trajectory according to this new set definition and repeat the procedure described above. Fig.\ \ref{ala-ran-4} shows the results for three correlation times (2 ps, 20 ps and 40 ps). It can be seen that the two sets of probabilities now deviate significantly at the shortest correlation times ($t_c$ = 2 and 20 ps), indicating that the new states are not as metastable as the old ones, and hence require significantly longer correlation times. As the correlation time is further increased to 40 ps however, the two probabilities nicely converge into an essentially perfect agreement. This again support the above discussion: while models built upon strongly metastable sets are preferable because $t_c$ can then be taken to be short compared to a typical first escape time from the set ($\sim 1/\lambda_1$), ultimate convergence is guaranteed for any set definition by simply increasing $t_c$.

\begin{figure}
\centering     
\subfigure[$\tau_c$=0.2 ns]{\label{vil-opt-4:a}\includegraphics[width=65mm]{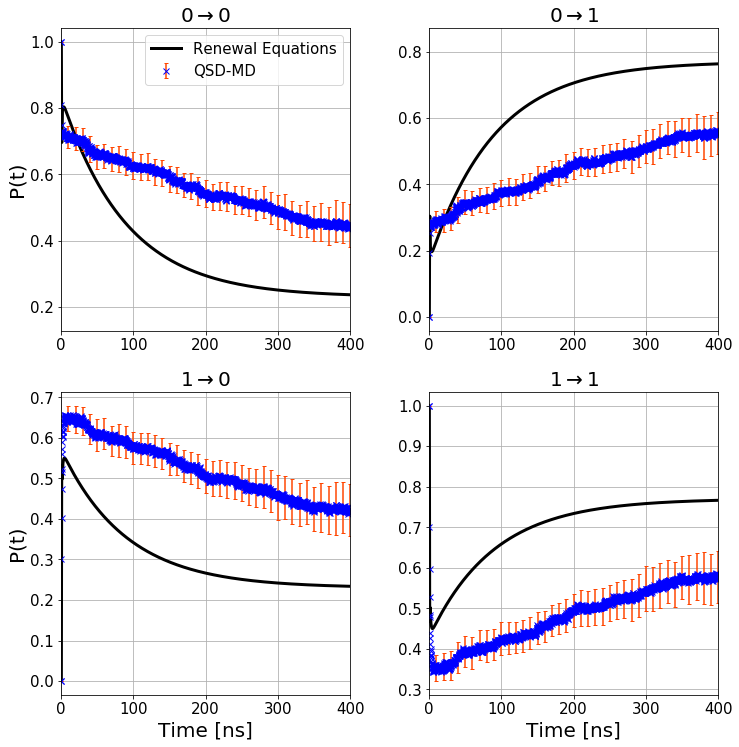}}
\subfigure[$\tau_c$=2 ns]{\label{vil-opt-4:b}\includegraphics[width=65mm]{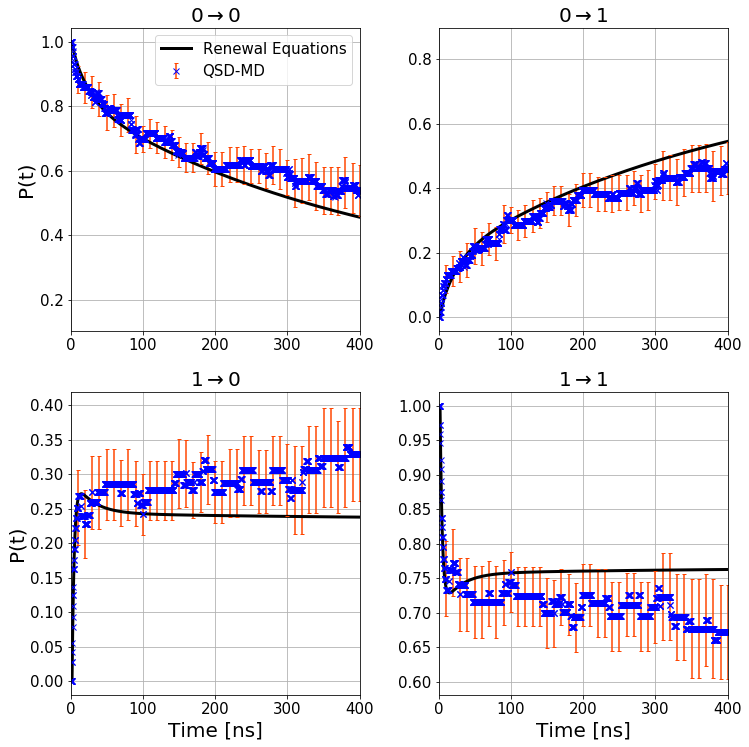}}
\subfigure[$\tau_c$=10 ns]{\label{vil-opt-4:c}\includegraphics[width=65mm]{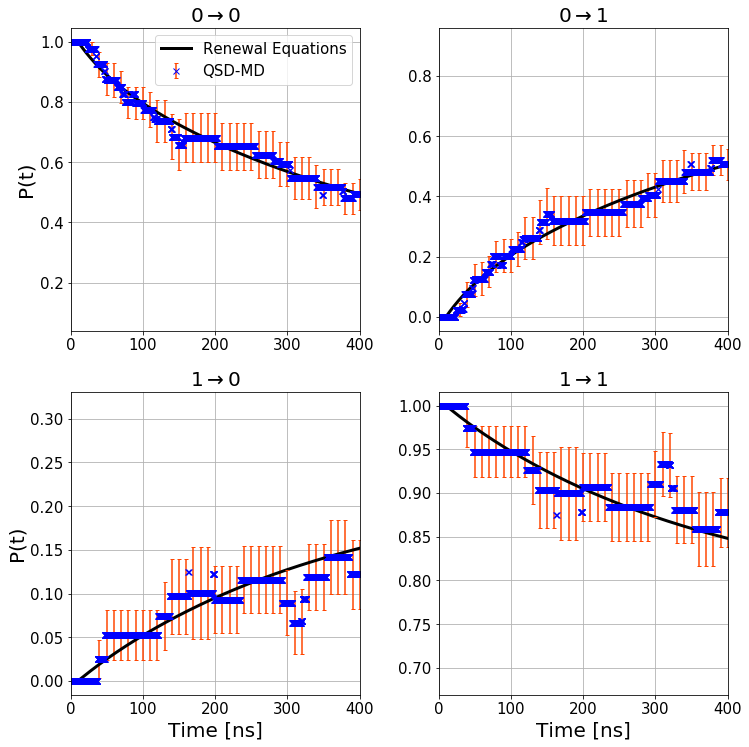}}
\caption{Transition probabilities computed from the discretized trajectory and the MRP model between the two PCCA
states of Chignolin. The error bars report the standard deviation obtained through a bootstrapping procedure where the trajectory data was subsampled before parameterizing the MRP. This subsampling was repeated 40 times.
}
\label{vil-opt-4}
\end{figure}

Fig.\ \ref{vil-opt-4} shows the same quantities for two PCCA states of Chignolin for three different correlation times ($t_c =$ 0.2 ns, 2 ns and 10 ns). For the shortest correlation time of 0.2 ns, the MRP  completely fails to capture the underlying dynamics. As the correlation time is increased, the transition probabilities computed from Eq.\ \ref{renewal} converge to the reference values, again showing that the approximation of the underlying dynamics can be systematically improved by increasing $t_c$. These encouraging results show promise that this formalism can be extended to more complex biological systems. Importantly, it can be valuable when atomistic information needs to be upscaled into coarser models \cite{di2019massively}.

\section{Conclusion}

We have shown that Langevin dynamics in a high-dimensional configuration space can be mapped to a jump process over a discrete state space through a combination of a novel mapping procedure, whereby the discrete state of the system corresponds to the last set in which the continuous trajectory spent a time $t_c$ without escaping, and of a MRP representation of the jump probabilities. This representation is shown to become exact as $t_c \rightarrow \infty$, and to converge exponentially fast to that limit, no matter the number and size of the sets used in the discretization, in contrast to conventional Markovian representations. We  therefore expect the MRP representation to accurately reproduce the dynamics of the system in a wide range of conditions, as supported by numerical examples. While this work demonstrates the formal power of this class of models, important questions related to efficient set definition and parametrization of MRPs from molecular dynamics simulations remain to be developed; these will be addressed in future publications.

\section{Acknowledgements}

We  thank D.E. Shaw Research for providing the chignolin trajectory.
A.A, S.G, and A.F.V acknowledge support from the Joint Design of Advanced Computing Solutions for Cancer (JDACS4C) program established by the U.S. Department of Energy (DOE) and the National Cancer Institute (NCI) of the National Institutes of Health.
D.P. ackowledges support from Los Alamos National Laboratory's (LANL) LDRD program under grant
20190034ER. LANL is operated by Triad National Security, LLC, for the National Nuclear Security Administration of U.S. Department of Energy (Contract No. 89233218CNA000001). Computing resources were made available by LANL's Institutional Computing program.

\bibliography{smm}{}

\end{document}